%% file: main.tex
\documentclass[cameraready]{Interspeech}

\usepackage{graphicx}
\usepackage{subcaption}
\usepackage{booktabs} 

\usepackage{hyperref}
\usepackage{amsmath}
\usepackage{amssymb}
\usepackage{mathtools}
\usepackage{amsthm}

\usepackage{multirow}
\usepackage[table]{xcolor}
\usepackage{lipsum}
\usepackage{listings}
\usepackage{verbatim}
\usepackage{tcolorbox}

\usepackage[capitalize,noabbrev]{cleveref}
\usepackage[
backend=biber,
style=ieee,
maxbibnames=3,
maxcitenames=3,
doi=false,isbn=false,url=false,eprint=false
]{biblatex} 
\defbibheading{bibliography}[\refname]{}

\theoremstyle{plain}

\theoremstyle{definition}

\theoremstyle{remark}

\newcommand{\real}{{\texttt{REAL}}}
\newcommand{\fake}{{\texttt{FAKE}}}

\newcommand{\tokenstart}{{\texttt{<TOF>}}}
\newcommand{\tokenend}{{\texttt{<EOF>}}}
\newcommand{\trnvocl}{{\texttt{Ft.Voc}}}
\newcommand{\trnll}{{\texttt{Ft.TTS}}}
\newcommand{\trnvl}{{\texttt{Ft.V+T}}}
\newcommand{\testvoc}{{\texttt{E.Voc}}}
\newcommand{\testpe}{{\texttt{E.PE}}}
\newcommand{\testav}{{\texttt{E.AV1M}}}

\newcommand{\testlm}{{\texttt{E.TTS}}}

\usepackage[textsize=tiny]{todonotes}

\title{Deepfake Word Detection by Next-token Prediction using Fine-tuned Whisper}

\author[affiliation={1}, equalcontribution]{Hoan My}{Tran}
\author[affiliation={2}, equalcontribution]{Xin}{Wang}
\author[affiliation={2}]{Wanying}{Ge}
\author[affiliation={2}]{Xuechen}{Liu}
\author[affiliation={2}]{Junichi}{Yamagishi}

\address{
    $^1$ Université de Rennes, France. $^2$ National Institute of Informatics, Japan
}
\email{hoan.tran@irisa.fr, wangxin@nii.ac.jp}


\begin{document}
\maketitle

\begin{abstract}
Deepfake speech utterances can be forged by replacing one or more words in a bona fide utterance with semantically different words synthesized with speech-generative models. While a dedicated synthetic word detector could be developed, we developed a cost-effective method that fine-tunes a pre-trained Whisper model to detect synthetic words while transcribing the input utterance via next-token prediction. We further investigate using partially vocoded utterances as the fine-tuning data, thus reducing the cost of data collection. Our experiments demonstrate that, on in-domain test data, the fine-tuned Whisper yields low synthetic-word detection error rates and transcription error rates. On out-of-domain test data with synthetic words produced with unseen speech-generative models, the fine-tuned Whisper remains on par with a dedicated ResNet-based detection model; however, the overall performance degradation calls for strategies to improve its generalization capability.
\end{abstract}

\section{Introduction}
\label{sec:intro}
Detecting deepfake speech utterances, which are not uttered by a real speaker but synthesized by a speech-generative model, is now a well-established topic with numerous publications on detector architectures~\cite{jung2022aasist,zhangAudio2024}, feature engineering~\cite{kambleAdvancesAntispoofingPerspective2020,takAutomatic2022}, datasets, and competitions~\cite{todiscoASVspoof2019Future2019,yiADD2023,kirillSAFE2025}. Beyond simply detecting whether an input utterance is synthetic (i.e., deepfake or forged with any synthesis model), an emerging task is to detect \emph{which part of the input utterance is synthetic}. Solutions are needed to counter advanced manipulations that alter part of a bona fide (i.e., human-uttered real) utterance, for example, by swapping bona fide words with synthetic ones.\footnote{As an official example of the VoiceCraft system by Meta, the original speech `governments don't control \emph{things}' was manipulated to `governments don't control \emph{money directly}'~\cite{peng2024voicecraft}.}Such manipulation can be conducted using traditional audio forgery methods (e.g., slicing) or more advanced deep-neural-network (DNN)-based generative models~\cite{zhangPartialEdit2025}. The latter case is known to produce synthetic content with less discernible artifacts~\cite{huangDetecting2024}.

Detecting the synthetic part of an utterance is more complicated than deciding whether the utterance is (partially) synthetic. For the latter, the detector has to produce a sequence of decisions (or scores that indicate the likelihood of being bona fide), rather than a single binary decision of synthetic or not. Accordingly, a sequential classification model can be implemented by sophisticating a binary deepfake speech detector, e.g., extracting sequential latent features from a DNN-based detector and transforming them via additional recurrent layers~\cite{zhang2022}, a boundary-matching map~\cite{chenLocalizing2025}, or attention mechanism~\cite{liFrameLevel2025}. 

In contrast to the above studies, we seek more cost-efficient solutions to locate the synthetic content in the input utterance. This is motivated by the fact that DNN-based models require a full-fledged development flow, including data curation, model designing, model training, and hyper-parameter tuning. Adding a trained model to a speech-data processing system also requires additional computation and storage resources. To avoid cost, we explore solutions that add synthetic word detection as an additional function to existing models, particularly automatic speech recognition (ASR) models. 

\begin{figure}
    \centering
    \includegraphics[width=\columnwidth, trim=0 360 0 8, clip]{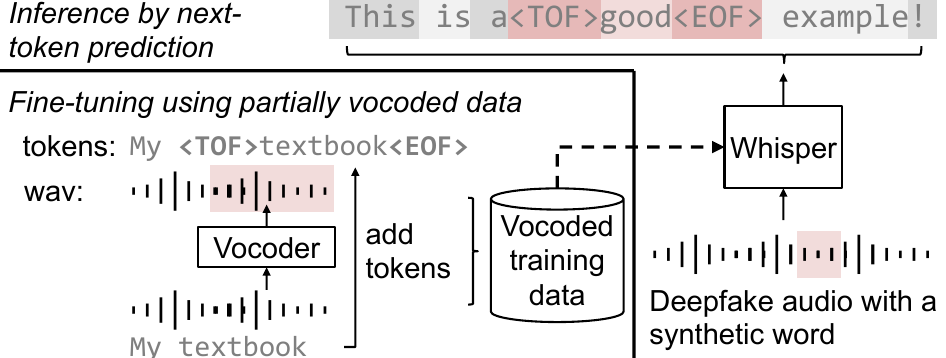}
    \vspace{-6mm}
    \caption{Illustration of fine-tuned Whisper for speech-to-text transcription and synthetic word detection. \tokenstart{} and \tokenend{} denote tokens surrounding synthetic word.}
    \label{fig:example}
    \vspace{-6mm}
\end{figure}

In this paper, we investigate \emph{whether we can fine-tune a pre-trained ASR model, i.e., Whisper~\cite{radfordRobust2022}, so that it can detect synthetic words while transcribing the input utterance.}\footnote{The term `synthetic word' refers to a  waveform segment that corresponds to a word, but its waveform signal is synthesized with a model. } 
To address this research question, we present a fine-tuning method that requires only a minimum change---adding specific tokens to the training data, which is illustrated in Figure~\ref{fig:example}. This avoids the costs of implementing a new training algorithm or deploying a separate fake detector. 
To reduce the cost of training-data curation, 
we also investigate the use of the training data that contains `simulated' synthetic words created with vocoders.

As far as the authors are aware, this is the first study to augment Whisper with the function of synthetic word detection. 
Our experiments demonstrate promising results. 
On test sets with data from the same domain as the training data, the fine-tuned Whisper performed on par with a dedicated ResNet model in terms of synthetic word detection, while its text transcription accuracy did not degrade. On test sets with words synthesized with advanced DNN-based generative models, the fine-tuned Whisper degraded to varying degrees in terms of synthetic word detection and transcription. Further analysis investigates how the fine-tuned Whisper is affected by the domain mismatch between fine-tuning and test data. Improving the out-of-domain detection performance is to be addressed in future work.

\section{Methods}
\subsection{Deepfake word detection via next-token prediction}
Let a speech waveform be $\boldsymbol{x}_{1:N}$ with $N$ denoting the total number of samples, where $x_n\in\mathbb{R}$ is the amplitude value at the $n$-th sampling point, $n\in[1, N]$. As an ASR model, Whisper uses an encoder-decoder Transformer structure and converts $\boldsymbol{x}_{1:N}$ into a sequence of $M$ tokens $\boldsymbol{y}_{1:M}$, wherein each token $y_m$ ($m \in \{1, \dots, M\}$) usually corresponds to a subword text unit and belongs to a finite set $\mathcal{Y}$ pre-defined by a tokenizer.

Conceptually, synthetic word detection involves converting $\boldsymbol{x}_{1:N}$ into another sequence of tags $\boldsymbol{c}_{1:M}$, where each $c_m\in\{\real{},\fake\}$. 
A na\"ive implementation is to add another classification head so that Whisper maps $\boldsymbol{x}_{1:N}\rightarrow \boldsymbol{y}_{1:M}\times\boldsymbol{c}_{1:M}$, but it requires a change in the model architecture and hyper-parameter tuning (e.g., weights for ASR and detection losses).

Our proposed method, illustrated in Figure~\ref{fig:example}, blends synthetic word detection and text transcription into \emph{a single next-token prediction task}. Fine-tuning Whisper only requires a minimal change to the tokens of the fine-tuning data. 
Suppose that we have pairs of speech utterances and their text transcriptions $\{\boldsymbol{x}_{1:N}, \boldsymbol{{y}}_{1:M} \}$, wherein each speech utterance contains one or more synthetic words. Assuming that the $i$-th token belongs to a synthetic word, we insert two tokens \tokenstart{} and \tokenend{} around $y_i$ to obtain a new token sequence $\boldsymbol{\hat{y}}_{1:M}=\{y_1, \cdots, \tokenstart{}, y_i, \tokenend{}, \cdots , y_M\}$. We now fine-tune Whisper using pairs of $\{\boldsymbol{x}_{1:N}, \boldsymbol{\hat{y}}_{1:M}\}$, without any change to the model architecture or fine-tuning algorithm. 
Inference is executed in the same way as the original Whisper. Given an output sequence of tokens, any token between a pair of \tokenstart{} and \tokenend{} is flagged as synthetic.

Note that, for $\{\tokenstart{}, \tokenend{}\}$, if we add two new tokens to $\mathcal{Y}$, we need to add additional embedding vectors to Whisper. We thus do not introduce new tokens but simply reuse two existing tokens from $\mathcal{Y}$ that are unlikely to be used for the ASR task. While other choices are possible, we use the tokens `\texttt{!!!!!!}' and `\texttt{$\sim\sim\sim$}' as the \tokenstart{} and \tokenend{}, respectively.

\subsection{Using vocoded data for fine-tuning Whisper}
\label{sec:method:vocoded}
A common approach to prepare the training data $\{\boldsymbol{x}_{1:N}, \boldsymbol{\hat{y}}_{1:M} \}$ is to use diverse speech-generative models to synthesize the selected words, which requires extra text or source audio input.
To simplify the data-curation pipeline, we are inspired by~\cite{wang2023spoofed} and use vocoders to create simulated synthetic words. Given a pair of $\{\boldsymbol{x}_{1:N}, \boldsymbol{y}_{1:M} \}$, we use WhisperX to obtain word-level alignment~\cite{bain23_interspeech}. We then randomly select one to five words, conduct copy-synthesis~\cite{wang2023spoofed} on their corresponding waveform segments via a vocoder, and replace the segments in the original waveform with the vocoded versions. This process only requires extracting the acoustic features (e.g., mel spectrogram) and resynthesizing the waveform. The vocoded waveforms are assumed to preserve the voice identity as well as the artifacts similar to those caused using speech-generative models~\cite{wang2023spoofed}.

Note that, when replacing the waveform segments with their corresponding vocoded counterparts, we use an overlap-add algorithm to ensure a smooth transition between the bona fide and the vocoded parts~\cite{zhang2022}. The source data and vocoders used in this study are explained in the next section.

\begin{table*}[t!]
    \centering
    \caption{Statistics of fine-tuning (\texttt{Ft.}) and test (\texttt{E.}) data sets. 
    Generators using full speech-generative systems are in italics.}
    \vspace{-3mm}
    \begin{tabular}{rrrrrl}
    \toprule
       & Name  & \#. of utterances & Language(s) & Domain & Generator of synthetic words  \\
    \midrule
    \multirow{3}{*}{\rotatebox{0}{\shortstack{Train \\ data}}}   
    & \trnvocl & 60,596 & en, es, fr, it, de & Audiobook & HiFi-GAN, NSF, NSF+GAN, WaveGlow, WORLD, GL  \\
    & \trnll & 60,596 & en & Audiobook & \emph{JETS, YourTTS, XTTS, SoVITS, CosyVoice, ElevenLab}  \\
    & \trnvl & 60,596 & en, es, fr, it, de & Audiobook & All above  \\
    \midrule
    \multirow{5}{*}{\rotatebox{0}{\shortstack{Test \\ data}}}
    & \testvoc & 3,000 & en, es, fr, it, de & Audiobook & HiFi-GAN, NSF, NSF+GAN, WaveGlow, WORLD, GL \\
    & \testlm  & 3,000 & en & Audiobook & \emph{JETS, YourTTS, XTTS, SoVITS, CosyVoice, ElevenLab}   \\
    \cmidrule{2-6}
    & \testav  & 3,000  & en & Youtube & \emph{YourTTS, VITS}  \\
    & \testpe  & 3,000  & en & Studio rec. & \emph{VoiceCraft, SSR-speech}  \\
    \bottomrule
    \end{tabular}
    \vspace{-4mm}
    \label{tab:data}
\end{table*}

\section{Experiments}
To answer the research question raised in \S~\ref{sec:intro}, 
we testified the models' performance on synthetic word detection as well as text transcription, using the training and test data listed in Table~\ref{tab:data}. 

\subsection{Data and protocols}
\textbf{Fine-tuning data}: we compared fine-tuning datasets using utterances altered by vocoders or speech synthesis systems. 
\begin{itemize}
    \item \texttt{\trnvocl{}}: a vocoded fine-tuning data set constructed on the basis of the MLS dataset~\cite{pratap20_interspeech}. We randomly selected 2,100 utterances per each of five languages: English (en), Italian (it), German (de), French (fr), and Spanish (es). In each utterance, one to five words were randomly selected and vocoded using the method in \S~\ref{sec:method:vocoded} and using each of six vocoders (or waveform reconstruction methods):  HiFi-GAN\cite{kong2020hifi}, WaveGlow~\cite{prenger2019waveglow}, Hn-NSF~\cite{wang2019neural}, a hybrid vocoder with Hn-NSF as the generator and HiFi-GAN's discriminators, WORLD~\cite{morise2016world}, and Griffin-Lim (GL)~\cite{griffinSignal1984}. The first four vocoders are based on DNNs, and the last two use signal-processing techniques. If an utterance failed to be vocoded (e.g., due to F0 extraction error), it was removed. This produced \trnvocl{} with around 60k utterances ($\approx 2100 \times 5 \times 6 $). 
    \item \texttt{\trnll}: a subset of the LlamaPartialSpoof corpus~\cite{luongLlamaPartialSpoof2025} with the same number of randomly selected utterances as \trnvocl{}. The synthetic words were generated using one of six text-to-speech (TTS) systems: JETS~\cite{limJETS2022}, YourTTS~\cite{casanova2022yourtts}, XTTS~\cite{casanova24_interspeech}, SoVITS~\cite{gpt-sovits}, CosyVoice~\cite{duCosyVoice2024a}, and ElevenLab\footnote{https://elevenlabs.io/}.
    \item\texttt{\trnvl}: a dataset with 50\% of the data randomly sampled from \trnvocl{} and the rest sampled from \trnll{}.
\end{itemize}
A validation set for \trnvocl{} was also created but using another 300 utterances per language. A validation set with the same size as that of \trnvocl{} was sampled from LlamaPartialSpoof for \trnll{}. The validation set for \trnvl{} was then merged from the two validation sets via random sampling. 

\textbf{Test data}: the \testvoc{} measures the performance on vocoded words, while the others measure the performance on data generated using DNN-based speech synthesis or editing systems. In terms of the data domain, \testvoc{} and \testlm{} are sourced from audio books, the same domain as the that of the training sets.
In contrast, \testav{} and \testpe{} are out-of-domain data since they are sourced from Youtube or a studio.
\begin{itemize}
    \item \texttt{\testvoc{}}: a vocoded data set constructed in the same procedure as \trnvocl{}. The utterances are disjoint from the training and validation sets. 
    \item \testlm{}: a subset of LlamaPartialSpoof with 3,000 utterances (disjoint from \trnll{} or \trnvl{}).
    \item \texttt{\testav}: a subset of the AV-Deepfake 1M validation set~\cite{caiAVDeepfake1M2024}. The synthetic words in the utterances were created using the TTS systems called VITS~\cite{kim2021conditional} and YourTTS~\cite{casanova2022yourtts}. 
    \item \texttt{\testpe{}}: a subset of  PartialEdit~\cite{zhangPartialEdit2025}, wherein synthetic words are created using publicly available attacks. 
    One is VoiceCraft~\cite{peng2024voicecraft}, an LLM-based speech-editing system that manipulates words via modifying the token. The other is also an LLM-based speech-editing system called SSR-speech~\cite{wangSSRSpeech2025}.
\end{itemize}

\subsection{Models and training configurations}

We used the pre-trained Large (v3) Whisper checkpoint~\cite{radfordRobust2022}. 
The fine-tuning was done on the entire model with a learning rate of 1e-5 and batch size of 8 (to fit a single Nvidia H100 GPU card).\footnote{We tried LoRA~\cite{huLoRA2022} but observed no improvement.} Training was conducted for five epochs, and the best checkpoint on the validation set was used for evaluation. 

We also trained a ResNet152~\cite{pindrop} for synthetic word detection, which was the sub-component of the top system in  the localization track of the 1M-Deepfakes Detection Challenge~\cite{pindrop}. The input waveform was transformed into mel spectrogram in the same setting as Whisper's front end. The model produces approximately one detection score per 16 ms. 
A training label (\real{} or \fake{}) was prepared per 16 ms given the ground-truth time alignment. 
Training was done using binary cross entropy for a maximum of 30 epochs, and the best checkpoint on the validation set was used for evaluation. 

\subsection{Evaluation metrics}

The text-transcription performance is measured using the word error rate (WER) implemented using the JIWER toolkit~\cite{jitsi_jiwer_nodate}. The WER is computed after removing $\{\tokenstart{}, \tokenend{}\}$ in the model's output and ground-truth transcription. 

Given the word alignment from the JIWER, the synthetic word detection is measured using the following metrics: 
\begin{itemize}
    \item false acceptance rate (FAR): among the synthetic words, the ratio of those being mis-classified as real,
    \item false rejection rate (FRR): among the real words, the ratio of those being mis-classified as synthetic.
\end{itemize}
Note that, even if a word is mis-recognized, it is not counted as a detection error when \real{} or \fake{} is correctly labeled. 
Because Whisper makes a hard decision when detecting the synthetic words, for a fair comparison, we let ResNet make a hard decision as well. For every word in a test utterance, the output probabilities from the ResNet's softmax layer aligned with the word were averaged as the probabilities of being \real{} (${P}_\text{real}$) and \fake{} (${P}_\text{fake}$). A word is classified as real if ${P}_\text{real} > {P}_\text{fake}$. 

\subsection{Results on in-domain data}
In \S~\ref{sec:exp:sdss}, we analyze the setups in which the fine-tuning and test data are matched in terms of data domain and synthesis methods, i.e., training on \trnvocl{}  and testing on \testvoc{} (\trnvocl{}  $\rightarrow$\testvoc{}), and training on \trnll{} and testing on \testlm{} (\trnll{}$\rightarrow$\testlm{}). In \S~\ref{sec:exp:sdds}, we discuss the crossed cases, i.e., \trnvocl{}$\rightarrow$\testlm{} and \trnll{}$\rightarrow$\testvoc{}. Evaluation across the domains is presented in \S~\ref{sec:exp:ood}.

\subsubsection{Matched data domain and synthetic methods}
\label{sec:exp:sdss}
On \testvoc{}, 
Whisper fine-tuned on \trnvocl{} obtained a lower WER (0.87\%) than the pre-trained Whisper (23.89\%).  
This is not surprising because both the fine-tuning and test data are in the domain of audiobooks. 
In terms of synthetic word detection, Whisper fine-tuned on \trnvocl{} obtained an FAR of 7.22\% and FRR of 0.52\%. The results are on a par with those of ResNet (FAR 7.15\% and FRR 3.81\%). 
We observed similar results on \testlm{}. Whisper fine-tuned on \trnll{} obtained a lower WER (2.20\%) than the pre-trained one (8.13\%). Meanwhile, the fine-tuned Whisper performed similarly (FAR 1.38\% and FRR 1.79\%) to ResNet in synthetic word detection (FAR 0.15\% and FRR 3.13\%). 

Note again that the training and test utterances do not overlap in terms of speech content. 
The results suggest that, \textbf{in the case where the data domain and synthetic methods match}, 
\textbf{the fine-tuned Whisper can well detect synthetic words while transcribing the input utterance via next-token prediction}. The fine-tuned Whisper performed on par with a well-known ResNet-based detection model, even though the detection is just an add-on to Whisper's transcription function. 

Example outputs of the fine-tuned Whisper are listed below, where wrong detection is marked in red. Note that the word \emph{Cay-Man} is correctly marked as \fake{}, even though it is not correctly transcribed. More examples are in supplementary materials.

\input{tab_error_1_v3}

\begin{tcolorbox}[left=1pt, right=1pt]
\footnotesize
\noindent Ground truth: $\qquad\quad$$\tokenstart{}$I$\tokenend{}$ $\tokenstart{}$present$\tokenend{}$ to you $\tokenstart{}$the$\tokenend{}$ human genome.

\noindent \trnvocl{}$\rightarrow$\testvoc{}: $\tokenstart{}$I$\tokenend{}$ $\tokenstart{}$present$\tokenend{}$ to you $\tokenstart{}$the$\tokenend{}$ human genome.

\noindent$\qquad$

\noindent Ground truth: But I could not bear the thought of $\tokenstart{}$wearing$\tokenend{}$ $\tokenstart{}$decayed$\tokenend{}$ $\tokenstart{}$boots.$\tokenend{}$

\noindent \trnll{}$\rightarrow$\testlm{}: But it could not bear the thought of $\tokenstart{}$wearing$\tokenend{}$ $\tokenstart{}$Cay-Man$\tokenend{}$ \textcolor{red}{boots}.

\end{tcolorbox} 

\begin{figure}[!t]
    \centering
    \subfloat[Error rates of fine-tuned Whisper on \testvoc{}]
    {
    \includegraphics[width=\linewidth]{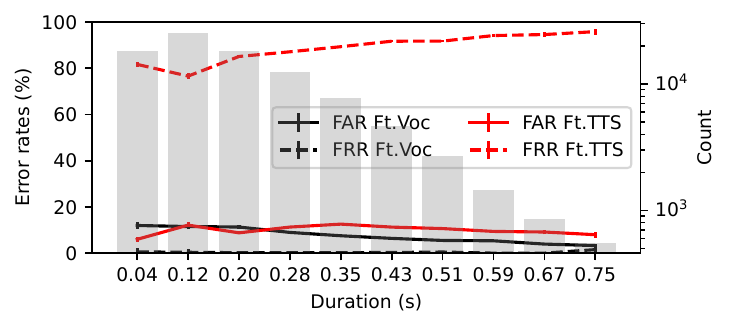}
    \vspace{-3mm}
    }\hfill
    \subfloat[Error rates of fine-tuned Whisper on \testlm{}]
    {
    \includegraphics[width=\linewidth]{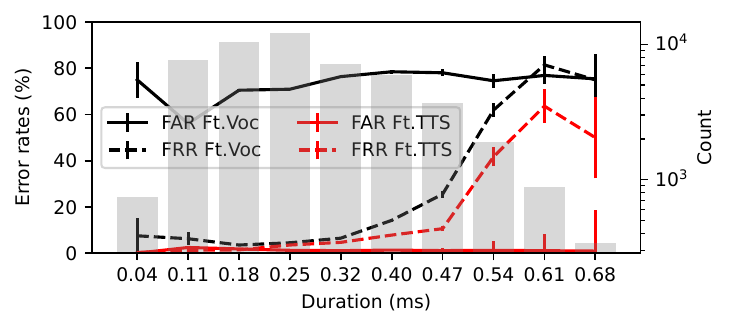}
    \vspace{-3mm}
    }
    \vspace{-3mm}
    \caption{Analysis of synthetic-word detection error rates based on word duration. Black and red profiles correspond to fine-tuned Whisper using \trnvocl{} and \trnll{}, respectively.}
    \vspace{-2mm}
    \label{fig:result:ana}
\end{figure}

\subsubsection{Matched data domain but different synthetic methods}
\label{sec:exp:sdds}

\textbf{When evaluating across the test sets in the same audiobook domain, the performance of fine-tuned Whisper and ResNet degraded to different degrees}. 

On \testvoc{}, the fine-tuned Whisper and ResNet trained from scratch using \trnll{} obtained higher FARs and FRRs than their counterparts using \trnvocl{}.
For analysis, we calculated the FARs and FRRs after grouping the words on the basis of their durations. Figure~\ref{fig:result:ana} (a) plots the results of Whisper fine-tuned on \trnvocl{} (black profiles) or \trnll{} (red profiles). 
While the FARs of the two fine-tuned Whispers are close, the FRR when using \trnll{} (red dashed line) is around or higher than 80\%. In contrast, the FRR when using \trnvocl{} (black dashed line) is close to 0\%. 

To further investigate the above results, we analyzed the errors across the languages using \testvoc{}. The results presented in Table~\ref{tab:result_vocoded} indicate that using \trnll{} led to FAR and FRR above 40\% on the English data. On other languages, it obtained FARs similar to those when using \trnvocl{} but FRRs close to 90\%. This suggests that the increased FRR when training on \trnll{} but testing on \testvoc{} is due to the unseen languages. 

On \testlm{}, the results in Table~\ref{tab:result_1} indicate that Whisper fine-tuned on \trnvocl{} obtained higher FAR and FRR than its counterparts using \trnll{}. Particularly, its FAR was as high as 76.16\%.  
As analysis, we plot the FARs and FRRs against the word duration in Figure~\ref{fig:result:ana}~(b) and observe that Whisper fine-tuned on \trnvocl{} obtained high FARs (black solid line) regardless of the word duration. 
A high FAR suggests that \textbf{Whisper fine-tuned on the vocoded training data struggles to spot words created using the TTS systems}. Using more diverse vocoders may improve performance, but exploration is left to future work.

Note that the FRR of Whisper fine-tuned on \trnvocl{} increases when the word is longer (black dashed line in Figure~\ref{fig:result:ana} (b)). 
The same trend is observed on Whisper fine-tuned using \trnll{} (red dashed line). 
One hypothesis is that the fine-tuned Whisper seems to flag \real{} only if it does not find any artifact within a word. The longer a word is, the more likely it contains patterns similar to artifacts.

\subsection{Results on out-of-domain test data}
\label{sec:exp:ood}
Given the above results, it is not surprising that the performance of the fine-tuned Whisper and ResNet also degraded on the out-of-domain test data. 
As Table~\ref{tab:result_2} shows, 
on both \testav{} and \testpe{}, we observe higher WER when Whisper is fine-tuned using either vocoded or TTS-synthesized data.  
In terms of synthetic word detection, the detection performance is overall unstable. For example, when training using \trnvocl{} and testing on \testpe{}, both models obtained low FRRs (~10\%) but very high FARs (~ 80\%).
In short, \textbf{current out-of-domain synthetic word detection is not yet sufficiently reliable.} 

Note that, when using \trnll{}, comparison between the results on \testlm{} and \testav{} shows that both the fine-tuned Whisper and ResNet performed worse on \testav{}, even though the synthesizers (\emph{YourTTS} and \emph{VITS}) were covered by the training data. Hence, to improve the out-of-domain performance, understanding the differences across datasets may be more important than complicating the model architecture.

\input{tab_error_lang}

\input{tab_error_2_v3}

\section{Conclusion}
We proposed a method for fine-tuning Whisper to detect synthetic words while preserving its speech-to-text transcription capability. The proposed method requires only augmenting the text token sequences with tokens that mark synthetic words, with no change to the algorithm or model architecture. This integrates the detection task into the next-token prediction task. 
In the experiment in which the training and test data matched, the fine-tuned Whisper, while executing transcription, detected the synthetic words with a performance comparable to a dedicated ResNet-based detection model. However, results on multiple test sets also indicate that the detection performance degraded when there was a data-domain mismatch between the training and test sets. A better understanding of the cross-domain differences is needed for future work. 
Using vocoded training data did not mitigate the degradation when the synthetic words were from a different domain or unseen synthesizers. Future work will investigate using vocoded data from diverse synthesizers and domains.  

\clearpage
\newpage

\section{Acknowledgment}
This work is partially supported by JST, PRESTO Grant (JPMJPR23P9), and K Program Grant (JPMJKP24C2), Japan. It is partially done on TSUBAME4.0, Institute of Science Tokyo.



\section{Generative AI Use Disclosure}
Generative AI was used to check grammatical errors.

\section{References}
{
\printbibliography
}

\newpage
\appendix

\onecolumn
\section{Example of synthetic word detection}
\label{sec:app:example}

\subsection{Example on vocoded test set \testvoc{}}
These are detection outputs using Whisper fine-tuned on \trnvocl{}. REF and HYP refers to the ground-truth and Whisper output. The start and end of a faked word are marked by `\texttt{!!!!!!}` and `$\sim\sim\sim$`, respectively.

Example 1: correctly identified all synthetic words 
\begin{lstlisting}[basicstyle=\ttfamily]
REF: !!!!!!Now~~~ this !!!!!!isn't~~~ absolutely definitive. 
     It's !!!!!!not~~~ to say that the idea isn't important.
HYP: !!!!!!Now~~~ this !!!!!!isn't~~~ absolutely definitive. 
     It's !!!!!!not~~~ to say that the idea isn't important.
\end{lstlisting}

Example 2: faked `weiter' is not detected (false acceptance), while real `leben' is incorrectly marked as fake (false rejection).
\begin{lstlisting}[basicstyle=\ttfamily]
REF: selber verdienen und        leben    !!!!!!weiter.~~~
HYP: selber verdienen und  !!!!!!leben~~~       weiter.
\end{lstlisting}

Example 2: faked `wie' and `und' are not detected (false acceptance), while real `dann' is incorrectly detected (false rejection).
\begin{lstlisting}[basicstyle=\ttfamily]
REF: solche Sachen !!!!!!wie~~~ !!!!!!dieses~~~ Licht !!!!!!und~~~
           dann    diese Baume, die
           
HYP: solche Sachen       wie    !!!!!!dieses~~~ Licht       und    
     !!!!!!dann~~~ diese Baume, die
\end{lstlisting}

\subsection{Example on AVDeepfake-1M test set \testav{}}
These are detection outputs using Whisper fine-tuned on \trnvocl{}.

Example 1: the synthetic word is corrected identified
\begin{lstlisting}[basicstyle=\ttfamily]
REF: To give kids that !!!!!!different~~~ experience and 
     hopefully them make those kind of friendships ...
HYP: To give kids that !!!!!!different~~~ experience and 
     hopefully them make those kind of friendships ...
\end{lstlisting}

Example 2: synthetic `want' and `it' are mis-classified as real (false acceptance) 
\begin{lstlisting}[basicstyle=\ttfamily]
REF: He !!!!!!doesn't~~~ !!!!!!want~~~ to talk about !!!!!!it~~~ 
     all the time, and whenever I start talking about it
HYP: He !!!!!!doesn't~~~       want    to talk about       it 
     all the time, and whenever I start talking about it
\end{lstlisting}

Example 3: `enjoy' is mis-recognized and mis-identified as a synthetic word (false rejection)
\begin{lstlisting}[basicstyle=\ttfamily]
REF: delivery so I really !!!!!!didn't~~~       enjoy     
     putting together my own album and finding my own sound.
HYP: delivery, so I really !!!!!!didn't~~~ !!!!!!kind~~~ of 
     putting together my own album and finding my own sound and
\end{lstlisting}

\subsection{Example on PartialEdit test set \testpe{}}
These are detection outputs using Whisper fine-tuned on \trnvocl{}.

Example 1: transcription error happened on `harmful', but the synthetic word is corrected marked.
\begin{lstlisting}[basicstyle=\ttfamily]
REF: Mixing drugs and alcohol can be extremely !!!!!!harmful.~~~
HYP: Mixing drugs and alcohol can be extremely !!!!!!humble.~~~
\end{lstlisting}

Example 2: the real `go' is incorrectly marked as fake (false rejection).
\begin{lstlisting}[basicstyle=\ttfamily]
REF: but you can       go    !!!!!!beneath~~~ that condition.
HYP: But you can !!!!!!go~~~ !!!!!!beneath~~~ that condition.
\end{lstlisting}

Example 3: the synthetic `fitting' is incorrectly marked as real (false acceptance).
\begin{lstlisting}[basicstyle=\ttfamily]
It seemed a moving and !!!!!!fitting~~~ !!!!!!destruction.~~~
It seemed a moving and       fitting    !!!!!!distraction.~~~
\end{lstlisting}


\end{document}

%% file: tab_error_1_v3.tex
\begin{table}[t!]
    \centering
    \setlength{\tabcolsep}{1.5pt}
    \caption{Results on \testvoc{} and \testlm{}. Bold text indicates the best result in each column and each test set. }
    \vspace{-3mm}
    \begin{tabular}{rrrrrrr}
    \toprule
        \multicolumn{2}{c}{Data set} & \multirow{2}{*}{\shortstack{WER(\%) \\ Whisper}} &
        \multicolumn{2}{c}{FAR (\%)} &
        \multicolumn{2}{c}{FRR (\%)} \\
       \cmidrule(lr){1-2}
       \cmidrule(lr){4-5}
       \cmidrule(lr){6-7}
       Test     &  Fine-tune      &     &  Whisper  &  ResNet  &  Whisper  &  ResNet \\ 
        \midrule
 \multirow{5}{*}{\rotatebox{90}{\testvoc}}
  & Pre-trained & 23.89 &       &       &       &        \\ 
  \cmidrule(l){2-7}
  & \trnvocl   &  \textbf{0.87} &  \textbf{7.22} &  \textbf{7.15} &  \textbf{0.52} &  \textbf{3.81} \\  
  & \trnvl &  2.18 &  8.45 &  9.80 &  1.01 &  5.32\\  
  &  \trnll    & 35.11 &  18.86 &  76.42 &  78.59 &  9.08\\
      \midrule
  \multirow{5}{*}{\rotatebox{90}{\testlm}}
  & Pre-trained &  8.13 &       &       &       &        \\ \cmidrule(l){2-7}
  &    \trnvocl{}     & 12.58 &  76.16 &  83.04 &  9.01 &  10.14\\
  &   \trnvl   & 3.86 &  2.08 &  0.34 &  3.82 &  32.90\\
  &    \trnll     & \textbf{2.20} &  \textbf{1.38} &  \textbf{0.15} &  \textbf{1.79} &  \textbf{3.13} \\  
        \bottomrule
    \end{tabular}
    \vspace{-3mm}
    \label{tab:result_1}
\end{table}

%% file: tab_error_lang.tex
\begin{table}[t!]
    \centering
    \caption{Decomposed results over languages using pre-trained Whisper (left) or fine-tuned on \trnvocl{} (right).}
    \vspace{-3mm}
    \begin{tabular}{rrrrr}
    \toprule
       & \multicolumn{2}{c}{FAR (\%)} & \multicolumn{2}{c}{FRR (\%)}
       \\ 
       \cmidrule(lr){2-3} \cmidrule(lr){4-5}
       &  \trnvocl{} &  \trnll{}  &  \trnvocl{}  &  \trnll{} \\ 
\midrule
en & \cellcolor[rgb]{0.98, 0.98, 0.98} 8.38 & \cellcolor[rgb]{0.85, 0.85, 0.85} 46.28 & \cellcolor[rgb]{1.00, 1.00, 1.00} 0.50 & \cellcolor[rgb]{0.87, 0.87, 0.87} 40.97\\ 
fr & \cellcolor[rgb]{0.97, 0.97, 0.97} 10.41 & \cellcolor[rgb]{0.97, 0.97, 0.97} 11.54 & \cellcolor[rgb]{1.00, 1.00, 1.00} 0.78 & \cellcolor[rgb]{0.59, 0.59, 0.59} 91.19\\ 
es & \cellcolor[rgb]{0.99, 0.99, 0.99} 5.59 & \cellcolor[rgb]{0.97, 0.97, 0.97} 10.76 & \cellcolor[rgb]{1.00, 1.00, 1.00} 0.36 & \cellcolor[rgb]{0.61, 0.61, 0.61} 89.36\\ 
de & \cellcolor[rgb]{0.99, 0.99, 0.99} 5.47 & \cellcolor[rgb]{0.98, 0.98, 0.98} 8.97 & \cellcolor[rgb]{1.00, 1.00, 1.00} 0.40 & \cellcolor[rgb]{0.59, 0.59, 0.59} 91.80\\ 
it & \cellcolor[rgb]{0.99, 0.99, 0.99} 6.30 & \cellcolor[rgb]{0.97, 0.97, 0.97} 13.03 & \cellcolor[rgb]{1.00, 1.00, 1.00} 0.53 & \cellcolor[rgb]{0.61, 0.61, 0.61} 88.12\\
    \bottomrule
    \end{tabular}
    \label{tab:result_vocoded}
    \vspace{-2mm}
\end{table}

%% file: tab_error_2_v3.tex
\begin{table}[t!]
    \centering
    \setlength{\tabcolsep}{2pt}
    \caption{Results on test sets manipulated using DNN-based speech synthesis and editing systems. Within each column, value with darker shading indicates worse performance.}
    \vspace{-3mm}
        \begin{tabular}{rrrrrrr}
    \toprule
        \multicolumn{2}{c}{Data set} & \multirow{2}{*}{\shortstack{WER(\%) \\ Whisper} } &
        \multicolumn{2}{c}{FAR (\%)} &
        \multicolumn{2}{c}{FRR (\%)} \\
       \cmidrule(lr){1-2}
       \cmidrule(lr){4-5}
       \cmidrule(lr){6-7}
       Test     &  Fine-tune      &     &  Whisper  &  ResNet  &  Whisper  &  ResNet \\
  \midrule
  \multirow{5}{*}{\rotatebox{90}{\testav}}
    & Pre-trained & \cellcolor[rgb]{1.00, 1.00, 1.00} 14.72  &       &       &       &         \\ 
  \cmidrule(l){2-7}
  &    \trnvocl     & \cellcolor[rgb]{0.91, 0.91, 0.91} 23.17 & \cellcolor[rgb]{0.80, 0.80, 0.80} 39.98 & \cellcolor[rgb]{0.84, 0.84, 0.84} 33.70 & \cellcolor[rgb]{0.99, 0.99, 0.99} 7.70 & \cellcolor[rgb]{0.94, 0.94, 0.94} 17.69\\ 
  &   \trnvl   & \cellcolor[rgb]{0.92, 0.92, 0.92} 21.23 & \cellcolor[rgb]{0.95, 0.95, 0.95} 15.98 & \cellcolor[rgb]{0.88, 0.88, 0.88} 27.63 & \cellcolor[rgb]{0.80, 0.80, 0.80} 39.42 & \cellcolor[rgb]{0.89, 0.89, 0.89} 27.04\\ 
  &    \trnll     & \cellcolor[rgb]{0.93, 0.93, 0.93} 20.47 & \cellcolor[rgb]{0.95, 0.95, 0.95} 16.04 & \cellcolor[rgb]{0.97, 0.97, 0.97} 11.14 & \cellcolor[rgb]{0.60, 0.60, 0.60} 59.97 & \cellcolor[rgb]{0.82, 0.82, 0.82} 36.44\\ 
  \midrule
  \multirow{5}{*}{\rotatebox{90}{\testpe}}
  & Pre-trained & \cellcolor[rgb]{1.00, 1.00, 1.00} 3.64 &       &       &       &          \\ 
  \cmidrule(l){2-7}
  &    \trnvocl     & \cellcolor[rgb]{1.00, 1.00, 1.00} 5.01 & \cellcolor[rgb]{0.66, 0.66, 0.66} 78.60 & \cellcolor[rgb]{0.59, 0.59, 0.59} 88.75 & \cellcolor[rgb]{0.99, 0.99, 0.99} 9.61 & \cellcolor[rgb]{0.98, 0.98, 0.98} 11.97\\
  &   \trnvl   & \cellcolor[rgb]{1.00, 1.00, 1.00} 4.81 & \cellcolor[rgb]{0.93, 0.93, 0.93} 28.39 & \cellcolor[rgb]{0.89, 0.89, 0.89} 36.56 & \cellcolor[rgb]{0.83, 0.83, 0.83} 49.90 & \cellcolor[rgb]{0.88, 0.88, 0.88} 40.44\\ 
  &    \trnll     & \cellcolor[rgb]{0.99, 0.99, 0.99} 5.77 & \cellcolor[rgb]{0.99, 0.99, 0.99} 8.74 & \cellcolor[rgb]{0.92, 0.92, 0.92} 30.92 & \cellcolor[rgb]{0.60, 0.60, 0.60} 87.89 & \cellcolor[rgb]{0.82, 0.82, 0.82} 52.49\\ 
        \bottomrule
    \end{tabular}
    \vspace{-2mm}
    \label{tab:result_2}
\end{table}